\documentclass[12pt]{article}

 \usepackage{amssymb}
 \usepackage{amsmath}
 \usepackage{amsfonts}
 \usepackage{color}
 \newcommand{\be}{\begin{equation}}
 \newcommand{\ee}{\end{equation}}
 \newcommand{\bea}{\begin{eqnarray}}
 \newcommand{\eea}{\end{eqnarray}}
 \newcommand{\nn}{\nonumber}
 \newcommand{\rd}{\partial}

\begin{document}

 \begin{titlepage}
  \thispagestyle{empty}

  \vspace{2cm}

  \begin{center}
    \font\titlerm=cmr10 scaled\magstep4
    \font\titlei=cmmi10 scaled\magstep4
    \font\titleis=cmmi7 scaled\magstep4
     \centerline{\titlerm  Near Horizon of 5D Rotating Black Holes}
     \centerline{\titlerm   from 2D Perspective  }

    \vspace{1.5cm}
    \noindent{{
        Hesam Soltanpanahi\footnote{e-mail: hesam@th.if.uj.edu.pl}
         }}\\
    \vspace{0.8cm}

   {\it Institute of Physics, Jagiellonian University, Reymonta 4, 30-059 Krak\'{o}w, Poland}

  \end{center}

  \vskip 2em

  \begin{abstract}
We study the CFT dual to five dimensional extremal rotating black holes, by investigating the two dimensional perspective of their near horizon geometry. From two dimensional point of view, we show that both gauge fields, related to the two rotations, appear in the same manner in the asymptotic symmetry and in the associated central charge. We find that,  our results are in perfect agreement  with the generalization of Kerr/CFT approach to five dimensional extremal rotating black holes.
  \end{abstract}

\end{titlepage}

 \tableofcontents

 \section{Introduction }\label{int}
 Five dimensional black holes have been interested from the seminal work on computing the entropy of a 5D black  hole by Strominger and Vafa in the context of string theory \cite{SV}. Recently by extending the Kerr/CFT \cite{GHSS} approach to 5D extremal rotating black  holes (ERBH) a wide class of such solutions have been studied \cite{LSB}-\cite{AS}. Common feature of these studies is the appearance of two rotating coordinates in the near horizon geometry of the most of these solutions as well as an AdS$_2$ part. It was proposed in some of these works that, there are two dual CFT's each of which corresponds to one of the rotations.

It was shown in \cite{LSK} that these two CFTs are related to each other by the SL(2,${\mathbb Z}$) transformation which is a symmetry in the space of the moduli parameters of the near horizon geometry of the 5D black  holes with two rotating coordinates. For this propose the boundary conditions  for the rotating coordinates are in the same order such that, the symmetry of the rotating coordinates is preserved by boundary fluctuations. This is not the case for the 5D black holes with only one rotation, e.g. black ring.

In this note we show the consistency of boundary conditions from 2D point of view  requires that the two gauge fields should be treated on the same footing in the study of asymptotic symmetry group. This leads to a chiral CFT, with central a special central charge, corresponding to the near horizon geometry of 5D extremal double rotating black  holes.

Each of the rotating coordinates in 5D reduces to one of the gauge field from 2D perspective. By introducing proper boundary conditions for the gauge fields and for the boundary energy-momentum tensor we calculate the corresponding central charge of the dual CFT. This approach  resembles the quantum entropy function introduced by Ashoke Sen \cite{S}, where all of the gauge fields are considered  in the same manner to study the thermodynamics of the extremal black holes.

By using the approach introduced by Castro and Larsen \cite{CL}, we reduce the near horizon geometry of 5D extremal double rotating black  holes to a 2D theory and investigate the properties of the boundary energy momentum tensor of the AdS$_2$ metric. We show that the variation of the energy momentum tensor under diffeomorphism which should be combined with gauge transformations \cite{HS} admits one central charge. As an example we calculate the associated central charge For Myers-Perry black  hole \cite{MP} and show the agreement with known results in this case.

The remaining of this paper is organized as follows. In \S\ref{5DERBH} we briefly review the 5D extremal rotating black hole and its CFT dual from 5D point of view. In \S\ref{2DP} following \cite{CL} we study the reduction of 5D extremal rotating black hole to the AdS solution of 2D theory. Then we derive the boundary terms of the 2D action and investigate the consistency of the boundary conditions which are allowed for 2D theory. By using the notion of Peierls bracket \cite{PB} and counter-term subtraction charge \cite{HIM} we define the associated charge and compute the central charge associated to the variation of the boundary energy momentum tensor  in \S\ref{CC}. In \S\ref{Examples} we study the Myers-Perry black  hole with two rotations and we show the agreement of our results with the previous calculations. Finally, \S\ref{CD} contains our conclusions and a brief discussion.
\section{ Review of 5D ERBH/CFT}\label{5DERBH}
This section is devoted to review the generalization of Kerr/CFT approach for 5D extremal rotating black  holes. The reader who is familiar with the Kerr/CFT approach can skip this section. We wil mention the main steps of calculations and will not discuss the details, which can be found in \cite{LSK}.

The near horizon geometry of 5D ERBH is given by \cite{KLR}\footnote{The AdS$_2$ radius has been absorbed in $F(\theta)$.}
\be
ds^2_5=F(\theta)ds^2_{AdS_2}+\sigma(\theta)d\theta^2+\gamma_{ij}(\theta)(dx^i+k^irdt)(dx^j+k^jrdt).\label{met0}
\ee
A possible boundary conditions for the fluctuations around this geometry (\ref{met0}) are \cite{LSK},
\be
h_{\mu\nu}\sim\mathcal{O}\left(\begin{array}{ccccc}
r^2&~ 1/r^2&~ 1/{r}&~ r &~ r \\
~&~{1}/{r^3}&~ {1}/{r^2}&~{1}/{r}&~{1}/{r}\\
~&~&~{1}/{r}&~{1}/{r}&~{1}/{r}\\
~&~&~&~{1}&~1\\
~&~&~&~&~1
\end{array}\right),\label{h}
\ee
in the basis $(t, r, \theta, \phi_1, \phi_2)$. These boundary conditions are consistent with the symmetry of the near horizon geometry which combine the $\phi_{1,2}$ coordinates with each other.

A general diffeomorphism preserving the boundary conditions (\ref{h}) is given by,
\bea
\zeta&=&\left[C+\mathcal{O}(\frac{1}{r^3})\right]\rd_t+\left[r\epsilon(\phi_1,\phi_2)+\mathcal{O}(1)\right]\rd_r+\mathcal{O}(\frac{1}{r})\rd_\theta\nn\\
&+&\left[\lambda_1(\phi_1,\phi_2)+\mathcal{O}(\frac{1}{r^2})\right]\rd_{\phi_1}+
\left[\lambda_2(\phi_1,\phi_2)+\mathcal{O}(\frac{1}{r^2})\right]\rd_{\phi_2},\label{solution}
\eea
where $\epsilon(\phi_1, \phi_2)$ and $\lambda_{1,2}(\phi_1, \phi_2)$  are arbitrary  smooth periodic functions of $\phi_1$ and $\phi_2$.

In \cite{LSK} it was shown that a class of diffeomorphism generators has  basis
\be
\zeta_m=-e^{-im\phi_1}\rd_{\phi_1}-e^{-im\phi_2}\rd_{\phi_2}-
2imr(e^{-im\phi_1}+e^{-im\phi_2})\rd_r,\label{zm}
\ee
and satisfy a Virasoro algebra $[\zeta_m , \zeta_n]_{\rm Lie}=-i(m-n)\zeta_{m+n}$. These generators correspond to a chiral CFT$_2$.

Using the definition of diffeomorphism charges \cite{BB, BC} and the Brown-Henneaux  approach \cite{BH}, it was shown that there is a Virasoro algebra between the associated charges with the central charge which is given by
 \be
 c=\frac{3(k_1+k_2)}{2\pi}\int d\theta
 d\phi_1d\phi_2\sqrt{\sigma(\theta)\gamma(\theta)}.\label{c-phi}
 \ee
We want to emphasize that, both of the $k_{1,2}$, which correspond to the angular momenta, contribute in the value of central charge in (\ref{c-phi}).

In the next two sections we confirm this result from the 2D perspective. For this propose we reduce the 5D near horizon geometry \eqref{met0} to 2D theory by integrating out the angular coordinates. The resulting solution has an AdS$_2$ metric and two gauge fields related to two angular momenta. We show that the combination of the diffeomorphism and gauge transformations of both of the gauge fields is consistent for investigating the variation of the boundary energy momentum tensor. In \S\ref{Examples} we show the agreement of 2D results with the 5D results for Myers-Perry black  holes.
\section{2D View of 5D Extremal Rotating Black Holes}\label{2DP}
In this section,  we want study the 5D ERBH from the 2D perspective. The next section is devoted to calculating the conserved charges and central charge following Castro and Larsen \cite{CL}. The steps and arguments are similar to \cite{CL} so we do not give all the details. By using the reduction we will show both of the gauge fields in 2D, associated to two rotating coordinates, play the same role in studying the asymptotic symmetry and the AdS/CFT correspondence. This is the 2D evidence for the arguments reviewed in \S\ref{5DERBH}.
\subsection{5D ERBH}
We start with the general form of the near horizon geometry of 5D ERBH with two angular momenta and by reduction on angular coordinates we obtain a 2D effective theory. As we mentioned in (\ref{met0}), the near horizon geometry of 5D ERBH with two rotations is given by
 \be
 ds^2_5=F(\theta)ds^2_{AdS_2}+\sigma(\theta)d\theta^2+\gamma_{ij}(\theta)(dx^i+k^irdt)(dx^j+k^jrdt),\label{met}
 \ee
where $F(\theta)$, $\sigma(\theta)$ and $\gamma_{ij}(\theta)$ are the functions of only $\theta$ and $x^i$ ,$i=1,2$ correspond to the rotating coordinates.

 By integrating out the the angular part, the two dimensional theory could be described by  a general 2D metric
 \be
 ds^2=g_{\mu\nu}dx^\mu dx^\nu,\label{gc1}
 \ee
 and two gauge fields corresponding to the rotations,
 \be
 \mathcal{A}^{i}=\mathcal{A}_\mu^{i}dx^\mu,\hspace{10mm}i=1,2,\label{gc2}
 \ee
 with $\mu,\nu=t,r$.
 We also couple the size of the angular coordinates to the scalar
 field $\psi$ such that
 \be
 ds^2=F(\theta)ds^2_2+e^{-2\psi}[\sigma(\theta)d\theta^2+\gamma_{ij}(\theta)(dx^i+\mathcal{A}^{i})(dx^j+\mathcal{A}^{j})].\label{met1}
 \ee
 Lowering and raising the indices  are defined by $g_{\mu\nu}$ and its inverse $g^{\mu\nu}$, respectively. The associated gauge field strengths are denoted by
 $\mathcal{F}^{i}=d\mathcal{A}^{i}$.

 The 5D Einstein Hilbert action is
 \be
 S_{(5)}=\frac{1}{16\pi G_5}\int d^5x\sqrt{-g_5}R^{(5)},\label{eha}
 \ee
 By using the  ansatz (\ref{met1}), one van find the the five dimensional Ricci scalar from two dimensional point of view,\footnote{We use indices,$\mu, \nu$ for coordinates $r, t$, and the indices $i, j$ for the gauge fields.}
 \bea
 R^{(5)}=\frac{1}{F(\theta)}\left[R^{(2)}-3e^{2\psi}\nabla^2e^{-2\psi}\right]
 -\frac{e^{-2\psi}}{F(\theta)^2}\gamma_{ij}(\theta)\mathcal{F}^{i}_{\mu\nu}F^{j,\mu\nu}+H(\theta)e^{2\psi}
 \eea
 in which
 \bea
 H(\theta)=\frac{1}{\sigma}&\bigg\{&\frac{1}{2}\left(\frac{d}{d\theta}\ln\frac{F}{\gamma}\right)^2
  -\frac{2}{F}\left(\frac{d^2F}{d\theta^2}\right)-\frac{3}{2\gamma}\textrm{det}|\frac{d}{d\theta}\gamma_{ij}|
 +\left(\frac{d}{d\theta}\ln F\gamma\right)\left(\frac{d}{d\theta}\ln\sigma\right)\nn\\
 &-&{1\over f}\left(\frac{d}{d\theta}\ln\sigma\right)
 \left[\gamma_{\phi\phi}\left(\frac{d^2\gamma_{\psi\psi}}{d\theta^2}\right)+
 \gamma_{\psi\psi}\left(\frac{d^2\gamma_{\phi\phi}}{d\theta^2}\right)
 -2\gamma_{\phi\psi}\left(\frac{d^2\gamma_{\phi\psi}}{d\theta^2}\right)\right]\bigg\},
 \eea
 and the five dimensional determinant is
 \be
 \sqrt{-g_5}=e^{-3\psi}\sqrt{F(\theta)^2\sigma(\theta)\gamma(\theta)}\sqrt{-g},
 \ee
 where
 \be
 \gamma(\theta)\equiv \textrm{det}|\gamma_{ij}(\theta)|.
 \ee

 By integrating over the angular coordinates, the 2D effective action of ERBH
 can be derived as
 \be
 S_{(2)}=\frac{\pi\alpha}{4G_5}\int d^2x\sqrt{-g}\left[ e^{-3\psi}R^{(2)}+\beta e^{-\psi}+
 {8\over3}\nabla_\mu e^{-\frac{3}{2}\psi}\nabla^\mu e^{-\frac{3}{2}\psi}
 - M_{ij}\mathcal{F}^{i}_{\mu\nu}F^{j,\mu\nu}e^{-5\psi}\right],\label{S2}
 \ee
 in which
 \bea
 &&\alpha=\int d\theta{\sqrt{\sigma(\theta)\gamma(\theta)}},\label{cons1}\\
 &&\beta={1\over\alpha}\int d\theta\sqrt{F(\theta)^2\sigma(\theta)\gamma(\theta)}H(\theta),\label{cons2}\\
 && M_{ij}={1\over\alpha}\int d\theta\frac{\sqrt{F(\theta)^2\sigma(\theta)\gamma(\theta)}}{F(\theta)^2}
 \gamma_{ij}(\theta).\label{cons3}
 \eea
The action (\ref{S2}) might be considered as the generic dilaton gravity in 2D with two gauge fields which was introduced in \cite{GKV}.

 \subsection{ Solutions}
 Since we are interested in solutions correspond to the geometry (\ref{met}),
 we limit ourselves to the solutions with constant $\psi$ and we try to solve the
 following  equations of motion,
 \bea
 3 R^{(2)}e^{-2\psi}-5 M_{ij}\mathcal{F}^{i}_{\mu\nu}F^{j,\mu\nu}e^{-4\psi}+\beta&=&0,\\
 \frac{1}{2}\left(\beta- M_{ij}\mathcal{F}^{i}_{\rho\tau}F^{j,\rho\tau}e^{-4\psi}\right)g_{\mu\nu}+
 2 M_{ij}\mathcal{F}^{i}_{\mu\rho}\mathcal{F}^{j,\rho}_{~~\nu}e^{-4\psi}&=&0,\\
 \nabla_\mu {F}^{i,\mu\nu}&=&0.
 \eea
 The first and second equations can be  simplified to
 \bea
  M_{ij}\mathcal{F}^{i}_{\mu\nu}F^{j,\mu\nu}=-\beta e^{4\psi}\\
 R^{(2)}=-2\beta e^{2\psi}
 \eea
 Assuming $\beta>0$, which is natural for reduction of extremal solution over angular coordinates,
 this solution is locally AdS$_2$ with radius
 \be
 l_{{\textrm AdS}}=(\frac{1}{\beta})^{1/2}e^{-\psi}\equiv
 le^{-\psi}.\label{l-b}
 \ee
 As one can see from (\ref{cons2}) $l_{AdS}$ is dimensionless.
 It is because we absorb the radius of the AdS$_2$ part of the near
 horizon geometry (\ref{met}) in $F(\theta)$.

 Without losing generality, we work in the gauge
 \be
 ds^2=e^{-2\psi}d\rho^2+g_{tt}dt^2,\hspace{10mm}\mathcal{A}^{i}_\mu
 dx^\mu=\mathcal{A}^{i}_t(\rho,t)dt.\label{gcgc}
 \ee
 In this gauge, the general form of the  solution of equations of motion are given by
 \bea
 g_{tt}&=&-\frac{1}{4}e^{-2\psi}\left(e^{\rho/l}-f(t)e^{-\rho/l}\right)^2\\
 \mathcal{A}^{i}_t&=&\frac{\lambda^i}{2l}e^{\rho/l}\left(1-\sqrt{f(t)}e^{-\rho/l}\right)^2\label{sol}
 \eea
 with the constraint
 \be
  M_{ij}\lambda^i\lambda^j=\frac{l^2}{2}.
 \ee
 Note that the constants $\lambda_i\neq0$ are inherited from $k_i\neq0$, which
 means the near horizon geometry of 5D ERBH (\ref{met}) has
 two non-zero angular momenta.

 It is convenient to  describe this solution in Fefferman-Graham coordinate expansion. The asymptotic behavior of the metric, scalar and gauge fields are given respectively by
 \bea
 g_{tt}^{(0)}&=&-\frac{1}{4}e^{-2\psi^{(0)}}e^{2\rho/l},\nn\\
 \mathcal{A}_t^{i(0)}&=&\frac{\lambda^i}{2l}e^{\rho/l},\label{ab}\\
 \psi^{(0)}&=&\textmd{costant}\nn
 \eea
 The result is similar to the reduction of four dimensional extremal Kerr black hole studied in  \cite{CL,CGLM}.

 \subsection{Boundary Terms}
 In this section, following the standard procedure for AdS/CFT we determine the
 normalized boundary action which is formally given by
 \be
 S_{\textmd{boundary}}=S_{\textmd{GHY}}+S_{\textmd{counter}}.
 \ee
 The first term is the Gibbons-Hawking-York term, namely
 \be
 S_{\textmd{GHY}}=\frac{2\pi\alpha}{4G_5}\int_{\rd\mathcal{M}} dt\sqrt{-h}~e^{{-3\psi}}K\
 \ee
 where $h$ and $K$ are respectively determinant of induced metric and  the extrinsic curvature  on the boundary $\rd\mathcal{M}$. It is easy to show that for solution (\ref{sol}) extrinsic curvature is
 \be
 K=\frac{1}{2}g^{tt}n^\mu\rd_\mu g_{tt}={1 \over l}e^\psi.\label{ec}
 \ee
 As discussed in \cite{CL,CGLM} the local form of the counter-term is given by
 \be
 S_{\textmd{counter}}=\frac{2\pi\alpha}{4G_5}\int_{\rd\mathcal{M}}
 dt\sqrt{-h}\left[m_1e^{{-2\psi}}+m_2e^{{-4\psi}}
 M_{ij}\mathcal{A}^{i}_{a}\mathcal{A}^{j,a}\right],
 \ee
 in which the constants $m_1$ and $m_2$ will be determined by vanishing the variation of the action on-shell.

 On the other hand, the variation of the action is given by
 \be
 \delta S=\int_{\rd\mathcal{M}}\left[\pi^{ab}\delta
 h_{ab}+\pi_\psi\delta\psi+\pi^a_{i}\delta\mathcal{A}_a^{i}\right]+\textmd{Bulk
 terms},\label{va}
 \ee
 with
 \bea
 \pi^{tt}&=&{\pi\alpha \over 4G_5}\left(m_1e^{-2\psi} h^{tt}+m_2e^{-4\psi}h^{tt} M_{ij}\mathcal{A}^{i}_{\mu}\mathcal{A}^{j,\mu}-
 2m_2e^{-4\psi} M_{ij}\mathcal{A}^{i,t}\mathcal{A}^{j,t}\right),\\
 \pi_\psi&=&{2\pi\alpha \over 4G_5}\left(-3e^{-3\psi}K-2m_1e^{-2\psi}-4m_2e^{-4\psi} M_{ij}\mathcal{A}^{i}_{a}\mathcal{A}^{j,a}\right),\\
 \pi^t_{i}&=&{\pi\alpha \over 4G_5}\left(-4e^{-5\psi} M_{ij}n_\mu\mathcal{F}^{j,\mu t}+4m_2e^{-3\psi}
 M_{ij}\mathcal{A}^{j,t}\right).
 \eea
 Using the asymptotic behavior of the fields (\ref{ab}) and the extrinsic curvature
 (\ref{ec}), the above expansions for conjugate momenta are reduced to
 \bea
 \pi^{tt}&=&{\pi\alpha \over 4G_5}\left(m_1+{m_2\over 2}\right)e^{-2\psi^{(0)}}h^{tt}_{(0)},\nn\\
 \pi_\psi&=&-{2\pi\alpha \over 4G_5}\left({3\over l}+2m_1-2m_2\right)e^{-2\psi^{(0)}},\label{pi}\\
 \pi^t_{i}&=&{2\pi\alpha \over 4G_5}\left(-1+m_2l\right) M_{ij}\lambda^j e^{-4\psi^{(0)}}h^{tt}_{(0)}e^{\rho/l}.\nn
 \eea
One can fix the constants $m_{1,2}$ by imposing vanishing boundary momenta \eqref{pi} which lead to  two conditions,
 \be
 m_1=-{1\over 2l},\hspace{10mm}m_2={1\over l}.
 \ee
 Note that, although there were two unknown constants $m_{1,2}$, we had three equations therefore, finding a solution shows the consistency of
 our calculations.
 In this way the full action of reduced solution is given by
 \bea
 S&=&\frac{\pi\alpha}{4G_5}\int_{\mathcal{M}} d^2x\sqrt{-g}\left[ e^{-3\psi}R^{(2)}+\beta e^{-\psi}+
 {4\over3}\nabla_\mu e^{-\frac{3}{2}\psi}\nabla^\mu e^{-\frac{3}{2}\psi}
 - M_{ij}\mathcal{F}^{i}_{\mu\nu}F^{j,\mu\nu}e^{-5\psi}\right]\nn\\
 &&+\frac{\pi\alpha}{2G_5}\int_{\rd\mathcal{M}}dt\sqrt{-h}\left[e^{-3\psi}K
 -{1\over2l}e^{-2\psi}+{1\over l}e^{-4\psi} M_{ij}\mathcal{A}^{i}_{a}\mathcal{A}^{j,a}\right]\label{fa}
 \eea
 \subsection{Consistency of Boundary Conditions}\label{CBC}
 As discussed in \cite{CGLM,HS}, for the AdS solution with a gauge field
 the combination of diffeomorphism and gauge transformation should be
 consistent with the gauge conditions.
 In this section we show that for solution (\ref{sol}), i.e. AdS metric with
 two gauge fields, the consistency requires that the gauge
 transformations of both of the gauge fields should be included in
 addition to the diffeomorphism.
 For this propose we first determine the diffeomorphism of
 the metric and its induction on the gauge fields. Then, we find the
 compensating gauge transformations leaving the gauge fields in the
 gauge condition (\ref{gcgc}).

 General diffeomorphism transforms the metric as
 \be
 \delta_\epsilon
 g_{\mu\nu}=\nabla_\mu\epsilon_\nu+\nabla_\nu\epsilon_\mu.
 \ee
 The gauge condition (\ref{gcgc}) has fixed the $g_{\rho\rho}$ and $g_{t\rho}$ components of the
 metric to zero and  by using the Fefferman-Graham form  we have fixed the asymptotic value of the $g_{tt}$ (\ref{ab}).
 Thus one can find the associated diffeomorphism which preserves these
 conditions by requiring the following conditions,
 \be
 \delta_\epsilon g_{\rho\rho}=0,\hspace{10mm}\delta_\epsilon
 g_{t\rho}=0,\hspace{10mm}\delta_\epsilon
 g_{tt}=0.\mathcal{O}(e^{2\rho/l}).\label{cbc2}
 \ee
 One can show that these conditions are satisfied if
 \be
 \epsilon^\rho=-l\rd_t\zeta(t),\hspace{10mm}\epsilon^t=\zeta(t)+2l^2\left(e^{2\rho/l}-f(t)\right)^{-1}\rd^2_t\zeta(t),\label{ad}
 \ee
 where $\zeta(t)$ is an arbitrary function of coordinate $t$.
 It is straight forward to find the transformation of the boundary metric under diffeomorphism  (\ref{ad}), namely,
 \be
 \delta_\epsilon
 h_{tt}=e^{-2\psi}\left(1-f(t)e^{-2\rho/l}\right)\left[{1\over2}\rd_tf(t)\zeta(t)+f(t)\rd_t\zeta(t)-l^2\rd^3_t\zeta(t)\right].
 \ee
 A general transformation of the gauge fields $\mathcal{A}^{i}_\mu$ defined by
 \be
 \delta_\epsilon\mathcal{A}^{i}_\mu=\epsilon^\lambda\nabla_\lambda\mathcal{A}^{i}_\mu+\mathcal{A}^{i}_\lambda\nabla_\mu\epsilon^\lambda,
 \ee
 also, leads to the following transformation
 under the same diffeomorphism,
 \be
 \delta_\epsilon\mathcal{A}^{i}_\rho=-2\lambda^ie^{-\rho/l}\left(1+\sqrt{f(t)}e^{-\rho/l}\right)^{-2}\rd^2_t\zeta(t).
 \ee
 To restore the gauge condition $\mathcal{A}^{(i)}_\rho=0$ (\ref{gcgc})
 one should compensate the diffeomorphism with a gauge
 transformations for each of the gauge fields as
 \be
 \mathcal{A}^{(i)}_\mu\rightarrow\mathcal{A}^{(i)}_\mu+\rd_\mu\Lambda^{(i)},\label{gts}
 \ee
 with gauge functions
 \be
 \Lambda^{i}=-2l\lambda^ie^{-\rho/l}\left(1+\sqrt{f(t)}e^{-\rho/l}\right)^{-1}\rd^2_t\zeta(t).
 \ee
 Therefore, the combination of the allowed diffeomorphism (\ref{ad}) and two gauge
 transformations (\ref{gts}) satisfy the gauge condition
 (\ref{gcgc}),\footnote{There is no summation over primed indices, in our notation.}
 \be
 \delta_{\epsilon+\Lambda^{i'}}\mathcal{A}^{i'}_\rho=\delta_\epsilon\mathcal{A}^{i}_\rho+\rd_\rho\Lambda^{i}_\rho=0.
 \ee
 One can easily show that under the combination of the
 transformations the variation of the gauge fields are
 \be
 \delta_{\epsilon+\Lambda^{i'}}\mathcal{A}^{i'}_t=
 {\lambda^i\over
 l}\bigg[e^{-\rho/l}\left({1\over2}\rd_tf(t)\zeta(t)+
 f(t)\rd_t\zeta(t)-l^2\rd^3_t\zeta(t)\right)-\rd_t\left(\zeta(t)\sqrt{f(t)}\right)\bigg].\label{gt}
 \ee

 Let us emphasize that, the gauge condition (\ref{gcgc}) can be satisfied if and only if
 the gauge transformations of both of the gauge fields
 (\ref{gt}) compensate with the the diffeomorphism (\ref{ad})
  and one can not turn off one of them consistently.

 From 5D point of view this means the rotating coordinates must
 play the same role in the asymptotic behavior of the metric.
 In other words, this implies that  the boundary conditions of rotating
 coordinates, which determine the fluctuations of associated
 components of the metric, should be in the same order.
 This is in precise agreement with the results of \cite{LSK} which is reviewed in \S\ref{5DERBH}.

 \section{Conserved Charges and Central Charge}\label{CC}

 Now we want to investigate the asymptotic symmetries by employing the associated conserved charges. Since we are interested in the boundary energy momentum tensor of a solution there are some subtleties to define the associated  conserved charges.

 As shown in \cite{HIM}, the generators of the asymptotic symmetries
 are  determined via the counter-term subtraction method (CTSM).
 These charges can differ from those defined usually.
 This method is based on the Peierls bracket \cite{PB} which has a
 covariant construction and is equivalent to the Poisson bracket on the
 space of observables.
 The charge calculated in this way is called counter-term subtraction
 charge (CTSC) which is given by
 \be
 Q_\xi=-\delta_{G,\xi}S,\label{dc}
 \ee
 where $\xi$ is an infinitesimal transformation parameter and $G$ is a regular function such that near the past boundary $G=0$
 and near the future boundary $G=1$.

 As we already mentioned, we focus on the boundary fields and the boundary energy
 momentum tensor.
 So, we need to determine the associated transformations.
 Using the induced transformation of an arbitrary boundary filed, $\Phi$ is defined by\footnote{ For details one can see \cite{HIM}.}
 \be
 \delta_{G,\xi}\Phi\equiv(\delta_{G\xi}-G\delta_\xi)\Phi.\label{dHIM}
 \ee
 we will study diffeomorphism and gauge transformation charges in the following subsections.

 \subsection{Diffeomorphism Charge}\label{DC}

 Under a general diffeomorphism transformation $x^\mu\rightarrow
 x^\mu+\epsilon^\mu$ the variation of the the full action is given
 by
 \be
 \delta_\epsilon S={1\over2}\int
 dt\sqrt{-h}T^{ab}\delta_\epsilon\gamma_{ab}+\int
 dt\sqrt{-h}J^a_i\delta_\epsilon\mathcal{A}_a^i+(\textmd{e.o.m})
 \ee
 where
 \bea
 T_{tt}&=&-{\pi\alpha\over 4G_5}\left({1\over l}e^{-2\psi}h_{tt}
 +{2\over l}e^{-4\psi} M_{ij}\mathcal{A}^{i}_{t}\mathcal{A}^{j}_t\right),\label{emt}\\
 J_{i,t}&=&{\pi\alpha\over G_5} M_{ij}\left(-n^\mu\mathcal{F}^{j}_{\mu t}e^{-\psi}
 +{1\over l}\mathcal{A}^{j}_{t}\right)e^{-4\psi},\label{cj}\\
 \delta_\epsilon\gamma_{ab}&=&\nabla_a\epsilon_b+\nabla_b\epsilon_a,\\
 \delta_\epsilon\mathcal{A}^i_a&=&\epsilon^b\nabla_b\mathcal{A}_a^i+\mathcal{A}_a^i\nabla_a\epsilon^b.
 \eea

 For induced transformation (\ref{dHIM}) the variation of the
 full action is simplified
 \bea
 \delta_{G,\epsilon} S&=&\int dt \sqrt{-h}T^{ab}\epsilon_a\nabla_bG
 +\int dt \sqrt{-h}J^a_i\mathcal{A}^i_b\epsilon^b\nabla_aG\nn\\
 &=&\sqrt{-h}h^{tt}(T_{tt}+J_{i,t}\mathcal{A}^i_t)\epsilon^t
 -\int dt\sqrt{-h}\nabla_a\left[(T^{ab}+J^a_i\mathcal{A}^{i,b})\epsilon_b\right]G\nn\\
 &=&\sqrt{-h}h^{tt}(T_{tt}+J_{i,t}\mathcal{A}^i_t)\epsilon^t~,\label{dge}
 \eea
  where in the last step we used the definition of Noether charge associated to the Peierls bracket.
  Using the definition CTSC (\ref{dc}) the associated charge is given by
  \be
  Q_\epsilon=-\sqrt{-h}h^{tt}(T_{tt}+J_{i,t}\mathcal{A}^i_t).
  \ee
  As we can see the energy momentum tensor and both of the gauge fields  appear in the diffeomorphism  generator.  One should note that, although the energy momentum tensor $T_{tt}$ and the $U(1)$ currents $J_{i,t}$ diverge as
  $\rho\rightarrow\infty$, the above combination is asymptotically finite, namely
  \be
  T_{tt}+J_{i,t}\mathcal{A}^i_t=
  {\pi\alpha\over4G_5}e^{-4\psi}\left({f(t)\over
  l}+\mathcal{O}(e^{-\rho/l})\right).
  \ee
  Note that for the extremal solution where $f(t)=0$  all
  non-extremal excitations will vanish.
  This is a consistency condition since it implies that
  the excitations considered above keeps the solution in the extremal limit \cite{GHSS}.

 \subsection{Gauge Transformation Charges}\label{GTC}
 Again one can explore  the variation of the action
 under a gauge transformation  $\delta_\Lambda\mathcal{A}^i_a=\rd_a\Lambda^i$
 by using the CTSM which is given by
 \bea
 \delta_{G,\Lambda^i}S&=&\int dt
 \sqrt{-h}\mathcal{J}_{i'}\Lambda^{i'}\rd_aG\nn\\
 &=&\sqrt{-h}\mathcal{J}^t_{i'}\Lambda^{i'}
 -\int dt\rd_a(\sqrt{-h}\mathcal{J}^t_{i'}\Lambda^{i'})G.\label{dlg}
 \eea
 where
 \be
 \mathcal{J}_{i}^a={\pi\alpha\over
 G_5l}e^{-4\psi}M_{ij}\mathcal{A}^{j,a}.
 \ee
 Similar to \eqref{dge}, the second term in (\ref{dlg}) vanishes, and by definition the charges of the
 gauge transformations are given by
 \be
 Q_{\Lambda^i}=-\sqrt{-h}\mathcal{J}^t_{i}=-{\pi\alpha\over
 G_5l}e^{-4\psi}\sqrt{-h}h^{tt} M_{ij}\mathcal{A}^j_t.
 \ee
 Using the asymptotic behavior of the gauge fields and metric (\ref{ab}) it is easy to show
 \be
 Q_{\Lambda^i}={\pi\alpha\over G_5l}e^{-3\psi}
 M_{ij}\lambda^j\left(1-2\sqrt{f(t)}e^{-\rho/l}+\mathcal{O}(e^{-2\rho/l})\right).
 \ee
 For the near horizon of extremal solutions the gauge transformation charges are
 given by
 \be
 Q_{\Lambda^i}={\pi\alpha\over G_5l} M_{ij}\lambda^j.
 \ee
 For the 4D extremal Kerr solution it was shown that this charge
 equals to the angular momentum in 4D point of view \cite{CL}.

 \subsection{Central charge}
 Now we can explore the combination of the physical generators. Moreover,
we can derive the central charge associated to the asymptotic transformations constructed in sections \S\ref{DC} and \S\ref{GTC}.

 The combined generator is given by
 \be
 Q_{(\epsilon+\Lambda^1+\Lambda^2)}=Q_\epsilon\epsilon+Q_{\Lambda^i}\Lambda^i,\label{QLL}
 \ee
 To study the transformations of this charge, it is natural to
 relate the transformation parameters, $\epsilon$ and $\Lambda^i$, to each other to treat the
 combined charge as a charge with one transformation parameter.
 The asymptotic behavior of the transformation parameters are given
 by
 \bea
 \epsilon^t&=&\zeta(t)+2l^2e^{-2\rho/l}\rd^2_t\zeta(t)+\dots,\\
 \Lambda^i&=&-2l\lambda^ie^{-\rho/l}\rd^2_t\zeta(t)+\dots,
 \eea
 and up to leading order one can write\footnote{There are some subtleties in this relation which are discussed in \cite{CL}. }
 \be
 \Lambda^i=l\mathcal{A}^i_a\rd_\rho\epsilon^a+\dots.\label{le}
 \ee
 Thus, the gauge transformations part of the combined charge (\ref{QLL})
become
 \be
 Q_{\Lambda^i}\Lambda^i=\sqrt{-h}h^{tt}\mathcal{A}_{t,i}\mathcal{J}^i_t\epsilon^t+\dots.\label{qle}
 \ee
 By employing equation (\ref{le}), it is easy to see the zeroth order of  $\epsilon^t$ does not appear in
  (\ref{qle}) and the first trem is $e^{-2\rho/l}$.

 Now we are able to study the transformations of the combined charge given in (\ref{QLL}).
 First we calculate transformation of the diffeomorphism part of the combined charge,
 $T_{tt}+J_{t,i}\mathcal{A}^i_t$, which is asymptotically given by
 \be
 \delta_{\epsilon+\Lambda^1+\Lambda^2}(T_{tt}+J_{t,i}\mathcal{A}^i_t)=
 2(T_{tt}+J_{t,i}\mathcal{A}^i_t)\rd_t\zeta(t)+\rd_t(T_{tt}+J_{t,i}\mathcal{A}^i_t)\zeta(t)
 +\mathcal{O}(e^{-\rho/l}).\label{defc}
 \ee
 On the other hand, the variation of the gauge transformations part of the combined charge (\ref{QLL}) are asymptotically
 \be
 \delta_{\epsilon+\Lambda^1+\Lambda^2}(\mathcal{J}_{t,i}\mathcal{A}^i_t)=
 \rd_t(\mathcal{J}_{t,i}\mathcal{A}^i_t)\zeta(t)
 -{\pi\alpha l e^{-4\psi}\over2G_5}\rd^3_t\zeta(t).\label{defc1}
 \ee
 It seems the weight of this part is zero
 but, as discussed in \cite{CL}, we do not worry about this fact.
 As mentioned above for this part of the combined charge, the asymptotic behavior is  $\epsilon^t\sim
 e^{-2\rho/l}$ and it has effectively weight two.

 Since the AdS$_2$ radius $l_{AdS}$ is dimensionless in the standard normalization of the transformation of the energy momentum tensor, the central charge is via
 \be
 \delta T_{ab}=2T_{ab}\rd_t\zeta(t)+\zeta(t)\rd_tT_{ab}
 -{c\over 12}\rd^3_t\zeta(t)~.\label{standard}
 \ee
Thus one can read the associated central charge by plugging (\ref{l-b}), (\ref{defc}), (\ref{defc1}) in (\ref{standard}), which is namely given by
 \be
 c={6\pi \alpha\over G_5\sqrt{\beta}}e^{-4\psi}~.\label{ccc}
 \ee
This is associated central charge for a chiral CFT dual to the AdS$_2$ geometry in the presence two gauge fields.

 \subsection{Levels}
 For completeness, we can find
the  level $k$ of the $U(1)$ gauge transformation which is defined by
 \be
 \delta_\Lambda J_t={k\over2}\rd_t\Lambda.
 \ee
 For the currents (\ref{cj}) associated to the gauge transformation of the $U(1)$
 charges one can derive the level by
 \bea
 \delta_{\Lambda^{i'}} J^{i'}_t
 ={\pi\alpha l\over G_5}e^{-4\psi}\rd_t\Lambda^{i}.
 \eea
 By using (\ref{l-b}) it is easy to show that the levels are determined by 
 \be
 k^{i}={\pi \alpha\over G_5\sqrt{\beta}}e^{-4\psi}.\label{kc}
 \ee
 Thus, both of the gauge fields have the same level, namely
 \be
 k^1=k^2=c/6.\label{kkc}
 \ee
 This relation between the central charge and the level of the
 R-currents is similar to the results of  \cite{CL,CGLM}.

 \section{Myers-Perry Black  Hole}\label{Examples}
 As an example to our results we study the Myers-Perry black  holes, which are simple 5D
 solution with two angular momenta in the near horizon geometry.
 The near horizon geometry of the Myers-Perry black  holes are of the
 form (\ref{met}). Without lose of generality we assume that $0<a<b$,
 where $a$ and $b$ are the two parameters of the Myers-Perry black  hole
 which are related to the two angular momenta. The
 parameters and functions of the metric (\ref{met}) for this solution are given by \cite{MP}
 \bea
 &&F(\theta)=\frac{\sigma(\theta)}{4},\hspace{7mm}\sigma(\theta)=ab+a^2\cos^2\theta+b^2\sin^2\theta,\\
 &&k^i\frac{\rd}{\rd x^i}=\frac{2\sqrt{a b}}{a(a+b)^2}\frac{\rd}{\rd\phi}+\frac{2\sqrt{a b}}{b(a+b)^2}\frac{\rd}{\rd\psi},\\
 &&f_{\phi\phi}(\theta)=\frac{(a+b)^2}{\sigma(\theta)^2}a\sin^2\theta(a+b\sin^2\theta),\\
 &&f_{\psi\psi}(\theta)=\frac{(a+b)^2}{\sigma(\theta)^2}b\cos^2\theta(b+a\cos^2\theta),\\
 &&f_{\phi\psi}(\theta)=\frac{(a+b)^2}{\sigma(\theta)^2}a b \sin^2\theta
 \cos^2\theta.\label{MPg}
 \eea

 Considering these expansions one can calculate the parameters
 (\ref{cons1}) and (\ref{cons2}) appeared in the central charge
 (\ref{ccc}) and the levels (\ref{kc}) as
 \bea
 \alpha=2(a+b)^2\sqrt{a b},\hspace{10mm}
 \beta={4a b\over(a+b)^2}.\label{MPab}
 \eea

 One can see from  (\ref{cons1}) that the parameter $\alpha$ is proportional to the
 area of the horizon which is essentially the Wald entropy of the black  hole.
 So it is natural to study the symmetries of the rotating coordinates
 which does not affect on the entropy. For  general study of this
 argument one can see \cite{LSK}.
 The geometry of Myers-Perry black  hole (\ref{MPg}) has a symmetry under $a\leftrightarrow b$
 compensated with $\theta\rightarrow\pi/2-\theta$.
 Thus after integrating over angular coordinates one would expect that the
 central charge (\ref{ccc}) has this symmetry.
 Using (\ref{ccc}), (\ref{kkc}) and (\ref{MPab})  the central charge and levels
 are given by
 \be
 c={3\over2}\pi(a+b)^3,\hspace{10mm}k_\phi=k_\psi={1\over4}\pi(a+b)^3
 \ee
 This is a sum of two central charges derived in \cite{LMP}.
 It is easy to show that, for Myers-Perry Black  Holes, this is in perfect agreement with the results obtained from five dimensional point of view reviewed in \S\ref{5DERBH} and is given in equation \eqref{c-phi}.

 \section{Conclusion and Discussion}\label{CD}
 In this paper we studied the near horizon geometry of the 5D extremal
 rotating black  holes from the 2D point of view by reducing the theory over
 the angular part of the coordinates following \cite{CL}.
 We showed that the consistency of the boundary conditions implies that
 both of the gauge fields, which correspond to two
 angular momenta in 5D, appear in the same manner.
 By studying the variation of the boundary energy momentum tensor we
 calculate the central charge of the CFT dual to the reduced
 solution which has a AdS$_2$ geometry.

 Although we did not trace the power of the fluctuations of the 5D metric
 due to the process of reduction,
 we showed that for consistent boundary conditions the results are
 in agreement with the calculations of 5D viewpoint \cite{LSK}.
 The advantage of the consistency is, compensating  of the
 variations of both of the gauge fields with the
 diffeomorphism variation studied in \S \ref{CBC}.
 It is interesting to study the relation between consistency of
 boundary conditions from 5D and 2D point of views which are given in (\ref{h}) and (\ref{cbc2}) respectively.

 Finally, we conclude that from a 2D point of view, the consistency of boundary
 conditions respect the symmetry of the near horizon geometry discussed in
 \cite{LSK} in the Kerr/CFT approach.
 Since parameter $\alpha$ appeared in the central charge (\ref{ccc}) is proportional to
 the Wald entropy (\ref{cons1}), it is natural to study the effect of the symmetry
 of the near horizon geometry.
 The symmetry of the rotating coordinates in 5D point of view
 inherited to $M_{ij}$ (\ref{cons3}) and its determinant.
 The result of this symmetry in 5D perspective was investigated in
 \cite{LSK} but, it is not realized in Myers-Perry Black  Holes studied in \S
 \ref{Examples}.

 The reduction of extremal black  holes in higher dimensions to 2D is
 also the basic argument of Sen's quantum entropy function \cite{S}.
 This reduction produce an AdS$_2$ metric, some scalar fields and a
 number of gauge fields associated to the angular momenta in higher dimensions.
 In \cite{S} it was shown that the thermodynamics of such solutions are determined 
 from quantum entropy function which is defined by
 \be
 d_{hor}(\overrightarrow{q})=\left<\exp[-iq_i\oint~d\theta
 A_\theta^{i}]\right>_{AdS_2}^{finite},
 \ee
 in the euclidean frame. Where $\oint~d\theta A_\theta^{i}$ denotes the integral of i-th
 gauge field along the boundary of AdS$_2$ and $q_i$ is the i-th
 electric charge.
 It is clear that all of the gauge fields play the same role in the
 thermodynamics of the reduced extremal solution.
 So, it is natural to expect that there is one CFT which is
 corresponded to the near horizon geometry of an ERBH.
 This was discussed for 5D ERBH in \cite{LSK} by employing the Kerr/CFT approach
 and  here we confirm the results from  2D point of view.
 It is worth to study  the relation between these approaches and Sen's
 quantum entropy function.

 Although we have not studied higher dimensional extremal black  holes we
 expect that one can apply this formalism in those case and after the reduction on angular coordinates,
 all of the gauge fields play the same role from 2D perspective and
 there is only one CFT corresponding to the near horizon geometry of extremal black  holes.

 Recently it was proposed a systematic  method for deriving the order
 of the boundary conditions of the metric for topologically massive gravity \cite{Sk}. Another interesting
 question is the extension of this method to higher dimensions and
 compare our results with this extension.

 In this paper we limited ourself to the solutions of 5D Einstein gravity but
 one can generalize this method to the solutions of other gravity theory e.g. supergravity.
 As a simple example with only one rotating coordinate one can
 study supersymmetric black ring \cite{EEMR}.
 Microscopic description of this solution is studied using other methods in  \cite{CGMS,ER,LSB}.

 \textbf{Acknowledgement}. I would like to thank Farhang Loran for useful discussions and comments on the draft and Department of Physics of Isfahan University of Technology where I did my PhD and part of this work was done. This work is supported by Foundation for Polish Science MPD Programm co-financed by the European Regional Development Fund, agreement NO. MPD/2009/6.


\end{document}